\documentclass[showpacs,preprintnumbers,amsmath,amssymb,APSl,prd,nofootinbib,superscriptaddress]{revtex4}
\usepackage{amsmath}
\usepackage{amssymb}
\usepackage{amsfonts}
\usepackage{graphicx,bm}
\usepackage{dcolumn}
\usepackage[colorlinks=true]{hyperref}
\usepackage{color,amsxtra}
\usepackage{epsf}
\usepackage{enumerate}
\usepackage{hhline}
\usepackage{array}
\usepackage{tabularx}
\usepackage{subfigure}
\usepackage{fancyhdr}
\usepackage{mathrsfs}

\newcommand{\be}{\begin{equation}}
\newcommand{\ee}{\end{equation}}
\newcommand{\bea}{\begin{eqnarray}}
\newcommand{\eea}{\end{eqnarray}}
\newcommand{\beaa}{\begin{eqnarray*}}
\newcommand{\eeaa}{\end{eqnarray*}}

\newcommand{\e}{\mathrm{e}}

\def\be{\begin{equation}}
\def\ee{\end{equation}}
\def\bea{\begin{eqnarray}}
\def\eea{\end{eqnarray}}

\begin{document}
\title{Testing the equation of state for viscous dark energy}
\author{Sergei D. Odintsov}
\email{odintsov@ice.csic.es} \affiliation{Institut de Ci\`{e}ncies de l'Espai,
ICE/CSIC-IEEC, Campus UAB, Carrer de Can Magrans s/n, 08193 Bellaterra (Barcelona),
Spain}

 \affiliation{Instituci\'o Catalana de Recerca i Estudis Avan\c{c}ats (ICREA),
Passeig Luis Companys, 23, 08010 Barcelona, Spain}
\affiliation{International Laboratory of Theoretical Cosmology, Tomsk State University of Control Systems and Radioelectronics (TUSUR) , 634050 Tomsk, Russia}
\affiliation{Institute of Physics, Kazan Federal University,  Kazan 420008, Russia}
\author{Diego S\'aez-Chill\'on G\'omez}
\email{diego.saez@uva.es} \affiliation{Department of Theoretical, Atomic and Optical Physics, Campus Miguel Delibes, \\ University of Valladolid UVA, Paseo Bel\'en, 7, 47011 Valladolid, Spain}
\author{German~S.~Sharov}
\email{sharov.gs@tversu.ru} \affiliation{Tver state university, Sadovyj per. 35, 170002 
Tver, Russia}

\begin{abstract}

Some cosmological scenarios with bulk viscosity for the dark energy fluid are considered. Based on some considerations related to hydrodynamics, two different equations of
state for dark energy are assumed, leading to power-law and logarithmic effective corrections to the pressure. The models are
tested with the latest astronomical data from Type Ia
supernovae (Pantheon sample), measurements of the Hubble parameter $H(z)$,
 Baryon Acoustic Oscillations and Cosmic Microwave Background
radiation. In comparison with $\Lambda$CDM model, some different results are obtained and their viability is discussed. The power-law model shows some modest results, achieved under
negative values of bulk viscosity, while the logarithmic scenario provide good fits in comparison to $\Lambda$CDM model.
\end{abstract}
%
%
\maketitle
%
%
%
\section{Introduction}

Over the last years, the intriguing question behind the late-time acceleration over the cosmological expansion has drawn much attention over the scientific community, being one of the most challenge problems in theoretical physics. As the universe is observed to be approximately homogeneous and isotropic at large scales, the best description for the cosmological evolution within General Relativity (and also within other geometrical theories) is provided by the so-called Friedmann-Lema\^itre-Robertson-Walker (FLRW) spacetime. Nevertheless, as can be easily shown trough the equations, an accelerating expansion in a FLRW universe requires generally an effective negative pressure fluid, which violates the energy conditions and has been called dark energy when talking on the late-time acceleration, although is also assumed for producing the so-called cosmic inflation at early times. Dark energy models have been deeply analysed, by considering each one's pros and cons, inherent of every theoretical model (see \cite{reviews}). However, the main remaining problem strikes on the similar predictions provided by a very large number of dark energy models, including new fields or modifications of General Relativity. Hence, a great effort is being done to reduce the number of models by studying more complex features, going beyond in perturbation theory or getting more accurate constraints by the use of the great amount of incoming data. \\

Moreover, as the main property for the dark energy fluid lies on the negativity of its pressure in order to achieve an accelerating expansion, a plausible scenario is provided by a viscous fluid, since its pressure is affected by a bulk viscosity term, as is well described in hydrodynamics, such that the fluid may keep the energy conditions satisfied, but providing an effective negative pressure. Such possibility has been widely analysed in the literature as well (for a review
see \cite{BrevikRev:2017}) and some realistic scenarios have been proposed where the well established knowledge of hydrodynamics is applied to cosmology \cite{Maartens}. In general, most of the analysis consider bulk viscosity as a possibility of assuming a viscous fluid, as can keep the conditions on homogeneity and isotropy, widely contrasted by the observations. In this sense, some authors have dealt with the possibility of incorporating viscosity to dark matter \cite{viscousDarkMatter}, which can lead to unify dark matter and dark energy under the same fluid, as for instance in the case of the Chaplygin gas \cite{Ferreira:2018knm,Chimento:2004bv,Lazkoz:2019ivd} or a logotropic fluid \cite{Ferreira:2016goc}. Some other models consider a proper dark energy fluid with viscosity \cite{CapozzielloCENO:2006,NojiriOInhEoS:2005,viscousDarkEnergy}. Such possibility may lead to a fluid with a negative pressure that may even cross the phantom barrier  \cite{PhantomViscous1,PhantomViscous2,NojiriOInhEoS:2005,CapozzielloCENO:2006,BrevikENO:2011,OdintsovOTSM:2018}. In addition, the viscosity terms may play an essential role during the early time inflationary stage \cite{BrevikRev:2017,Gron:1990}.


In the present paper we consider two models for dark energy with bulk viscosity, in the same line as proposed in some works previously. Here our models are based on some phenomenological considerations or inspired by condensed matter physics. In that sense, we study a model whose viscosity depends on the powers of the energy density and the Hubble parameter, being considered as effective corrections to a perfect dark energy fluid, while the second model is inspired in Anton-Schmidt's equation of state for crystalline solids (\cite{viscousLog}). We test the model by using some recent observational
data, and techniques developed in some previous papers
 \cite{OdintsovSGS:2017,Odintsov_Flog:2019,Sharov16,PanSh17}. The observational datasets
include the latest Pantheon sample \cite{Pantheon17} of Type Ia supernovae (SNe Ia),
estimations of the Hubble parameter $H(z)$, observational manifestations of baryon
acoustic oscillations (BAO) and cosmic microwave background radiation (CMB)
\cite{Eisen05,Planck13,Planck15,Planck18}. By using the likelihood, we obtain the best fits for the free parameters of the models and compare to $\Lambda$CDM model.\\

The paper is organized 
 as follows: section \ref{Dynamics} is devoted to a brief description of dynamical equations and the models for the bulk viscosity.
 In Section \ref{Data}, we provide the observational
datasets used along the paper for testing the models, which correspond to SNe Ia, $H(z)$, BAO and CMB under investigation. Section \ref{Results} is
devoted to the results of the analysis of the models. Finally, in Section \ref{Conclusion} we summarise the
results of this work.

\section{Background}
\label{Dynamics}

Let us start by introducing the basis of the paper. Here we are assuming a flat Friedmann-Lema\^itre-Robertson-Walker (FLRW) metric, which can be expressed in co-moving coordinates as follows:
 \be
 ds^2 = -dt^2+a^2(t)\Big[(1-k r^2)^{-1}dr^2+r^2 d\Omega\Big]\ ,
\label{FLRWmetric}
 \ee
Here we are interested in the analysis of viscous fluids, such that the bulk viscosity $\zeta$ is introduced as an effective contribution to the pressure \cite{BrevikRev:2017,NojiriOInhEoS:2005,CapozzielloCENO:2006}:
 \begin{equation}
p =w(\rho)\cdot\rho + B(H),\qquad B(H)=- 3H\zeta(H).
 \label{EoSvisc}
\end{equation}
Note that $\zeta(H)$ depends on time and can be written in terms of the Hubble rate $H = \dot{a}/a$ as we are assuming a FLRW spacetime. Hence, the effective pressure $p$ includes the viscous term $B(H)$, which would satisfy the transportation equation, and may be extended to more general forms that contain derivatives of the Hubble parameter $B=B(a,H,\dot
H)$ (see Ref.~\cite{BrevikRev:2017,NojiriOInhEoS:2005,CapozzielloCENO:2006}). Here, we are considering two known models for the bulk viscosity (\ref{EoSvisc}). \\

In the first scenario, we also assume a non constant equation of state but depending on
powers of the energy density $w=w(\rho)$ while the viscosity is given by $\zeta(H)\sim
H^{2\beta-1}$. Then, the effective EoS yields
\cite{NojiriOInhEoS:2005,CapozzielloCENO:2006,BrevikENO:2011}:
 \begin{equation}
p =-\rho +A \rho^{\alpha} + BH^{2\beta}\ .
 \label{EoS}
\end{equation}
The second model is inspired in Anton-Schmidt's equation of state for crystalline solids (\cite{viscousLog}), which includes a logarithmic-corrected power-law fluid and the
the same viscous term as above $\zeta(H)\sim H^{2\beta-1}$, leading to the following EoS \cite{OdintsovOTSM:2018}:
 \begin{equation}
p = A\bigg(\frac\rho{\rho_*}\bigg)^{\alpha} 
\log\bigg(\frac\rho{\rho_*}\bigg) + BH^{2\beta}\ .
 \label{EoSln}
\end{equation}
For both models, $A$, $\alpha$, $ B$, $\beta$, $\rho_*$ are constants, essentially the free
parameters of the model, while $\alpha=\gamma_G+\frac16$,  where  $\gamma_G$ is
the Gr\"{u}neisen parameter. The density $\rho_*$ shows the limit where standard pressure vanishes, and can be identified with the
Planck density $\rho_P = c^5/(\hbar G)$ (see Ref.~\cite{Chavanis:2015,CapozzielloAL:2018}. For aesthetic, we assume the units such as the speed of light reduces to unit $c=1$. \\

In the sections below, we will test the viability of models (\ref{EoS}) and (\ref{EoSln}) by confronting their predictions with recent observational
data, coming from different sources. Besides the dark energy component $\rho_x$, we will also take into account the other two components that play an important role along the cosmological evolution, dust (baryons and cold dark matter) $\rho_m$ and radiation $\rho_r$, such that the total energy density can be expressed as follows:
  \begin{equation}
\rho = \rho_m+\rho_x+\rho_r
 \label{rhotot}
\end{equation}
 In addition, we also assume that there is no interaction among the three components, such that they satisfy the continuity equation independently,
\begin{equation}
\dot{\rho}_i+ 3 H (p_i+\rho_i)= 0.
 \label{cont} \end{equation}
Let us now consider the scenario with the three matter components (\ref{rhotot}) with energy densities
$\rho_m$, $\rho_x$, $\rho_r$, where $\rho_x$ is governed by the EoS given in
(\ref{EoS}) or (\ref{EoSln}). In particular, by the power-law
EoS (\ref{EoS}), the expression for the pressure leads to $p_x =-\rho_x +\tilde A\rho_x^{\,\alpha} + \tilde BH^{2\beta}.$ As usual in a FLRW Universe, cold dark matter $\rho_m$ and radiation
$\rho_r$  evolve according to their continuity equations respectively (\ref{cont}):
 \begin{equation}
  \rho_m=\rho_m^0a^{-3},\qquad \rho_r=\rho_r^0a^{-4}.
 \label{rhomr}\end{equation}
Here the index $0$ refers to magnitudes measured at the present time $t_0$ while the scale factor at the present time is set as the unity, $a(t_0)=1$.
In this notation, the redshift $z$ of a luminous object is $z=a^{-1}-1$. On the other hand, it is convenient to rewrite the equations of state (\ref{EoS}) and (\ref{EoSln}) for the
viscous component in the following form
 \begin{eqnarray}
\frac{p_x}{\rho_{\mbox{\scriptsize cr}}} &=&-\Omega_x +A\Omega_x^{\,\alpha} +
B\bigg(\frac{H}{H_0}\bigg)^{2\beta}, \quad(\mbox{``power-law'' model}), \label{EoS1}\\
\frac{p_x}{\rho_{\mbox{\scriptsize cr}}} &=&A\Omega_x 
\log\frac{\Omega_x}{\Omega_*} + B\bigg(\frac{H}{H_0}\bigg)^{2\beta},
\quad\;\;(\mbox{``logarithmic'' model}); \label{EoSln1}
\end{eqnarray}
 where the dimensionless cosmological parameter $\Omega_x$ is defined as usual by the ratio among the dark energy density and the critical density:
\be
\Omega_x=\frac{\rho_x}{\rho_{\mbox{\scriptsize cr}}}=\frac{\kappa^2\rho_x}{3H_0^2}\ , \quad \Omega_*=\frac{\rho_*}{\rho_{\mbox{\scriptsize cr}}}\ ,
\label{CosmoParametDE}
\ee
where the Hubble constant  and the critical density are given by $H_0=H(t_0)$ and $\rho_{\mbox{\scriptsize cr}}=3H_0^2\big/\kappa^2$, respectively, while the constant $\kappa^2\equiv8\pi G$.  In the logarithmic model (\ref{EoSln1}) we consider only the case $\alpha=1$ because of too large number of model parameters.\\

Then, by using the Einstein field equations, together with the FLRW metric (\ref{FLRWmetric}), the corresponding FLRW equations are obtained:
\bea
H^2+\frac k{a^2}&=&\frac{\kappa^2}3\rho,  \label{Fried} \\
2 \frac{\ddot a}a+H^2+\frac k{a^2}&=&-\kappa^2p\ ,
 \label{Fried2} \eea
where $\rho=\sum\rho_i$  corresponds to (\ref{rhotot}). Recall that the continuity equation (\ref{cont}) can be retrieved by combining (\ref{Fried}) and (\ref{Fried2}), such that is not an independent equation, as usual in covariant theories. We can rewrite the FLRW equations (\ref{Fried}) and (\ref{Fried2}) in terms of the cosmological parameters (\ref{CosmoParametDE}) as follows:
\begin{equation}
 \frac{H^2}{H_0^2}=\Omega_m^0 a^{-3}+\Omega_x(a)+\Omega_r^0 a^{-4}+\Omega_ka^{-2}, \label{EqF}\ ,
 \end{equation}
whereas the continuity equation (\ref{cont}) can be expressed as function of the scale factor instead of the cosmic time for the models (\ref{EoS1}) and (\ref{EoSln1}):
\begin{equation}
\frac{d\Omega_x}{d \ln a} 
=\left\{\begin{array}{ll}
 -3\Big[A\Omega_x^{\,\alpha} +
B\big({H}/{H_0}\big)^{2\beta}\Big], & \mbox{(power-law)},\\
-3\Big[\Omega_x+A\Omega_x 
\log \frac{\Omega_x}{\Omega_*} +
B\big({H}/{H_0}\big)^{2\beta}\Big],\;\; & \mbox{(logarithmic)}. \rule{0mm}{1.5em}
 \end{array}\right. \label{Eqcont}
 \end{equation}
Here
 $$\Omega_m^0=\frac{\rho_m^0}{\rho_{\mbox{\scriptsize cr}}},\qquad
 \Omega_r^0=\frac{\rho_r^0}{\rho_{\mbox{\scriptsize cr}}},\qquad \Omega_k=-\frac{k}{H_0^2}.$$
Hence, by solving the system of equations (\ref{EqF}) and (\ref{Eqcont}), the cosmological evolution is obtained in terms of the scale factor for some values of the free parameters, together with the corresponding expressions for radiation and dust in terms of the scale factor, given in Eq.~(\ref{rhomr}). \\

In order to simplify the model and reduce the number of free parameters, we are considering a flat FLRW universe, such that the curvature is assumed to be zero and the corresponding cosmological parameter leads to $\Omega_k=0$. Moreover, we will also fix the cosmological parameter for radiation density $\Omega_r^0$ through the ratio among baryons and radiation as provided by Planck \cite{Planck13}:
\begin{equation}
 X_r=\frac{\rho_r^0}{\rho_m^0}=\frac{\Omega_r^0}{\Omega_m^0}=2.9656\cdot10^{-4}\,.
  \label{Xrm}\end{equation}
Since this value is rather small, the radiation density $\rho_r$  is assumed as negligible when fitting our models with SNe Ia,
$H(z)$ and BAO observations in the range $0<z\le2.36$, as usual in most of the analysis of this kind. The component $\rho_r$ becomes important just at high redshifts, at which radiation density turns out essential to deal with is the corresponding CMB observational data for redshifts $z\simeq1000$.

Hence, by setting the spatial curvature to be zero $\Omega_k=0$ and by fixing the radiation-matter ratio
(\ref{Xrm}), the free parameters for both models
(\ref{EoS1}) and (\ref{EoSln1}) turn out:
 \begin{equation} \begin{array}{ll}
\Omega_m^0, \; A,\;\alpha,\; B,\;\beta,\; H_0, & \mbox{ \ (power-law model)},\\
\Omega_m^0, \; A,\; B,\;\beta,\;\Omega_*,\; H_0, \;\;& \mbox{ \ (logarithmic model)}.
  \rule{0mm}{1.2em}\end{array}
 \label{6param}
\end{equation}
Note that the large number of parameters $N_p$  is a lack of strength for any model in comparison with other
cosmological scenarios, as the $\Lambda$CDM model, since the increasing number of free parameter may lead to a loss of information and to weaker constraints on the free parameters, which may flag the corresponding theoretical model from the point of
view of information criteria \cite{Akaike74,Schwarz78}. However, we
 consider the Hubble constant $H_0$ as a nuisance parameter and reduce the
effective number to $N_p=5$. In addition, we will also show that the
power-law model (\ref{EoS1}) with $\alpha=1$, provides similar fits and errors as the case when $\alpha$ is considered as a free parameter, which together with the low correlation among the parameters, gives reliable results and constraints on the models.

\section{Observational data}
\label{Data}

Let us introduce now the data that will be used to test and comapre the models
(\ref{EoS1}) and (\ref{EoSln1}). These sets of data include the largest recent catalogue
of Type Ia supernovae (SNe Ia), the so-called Pantheon sample \cite{Pantheon17}, baryon
acoustic oscillations (BAO) data \cite{Eisen05,BAOdata}, estimations of the Hubble
parameter $H(z)$ \cite{Hdata} and parameters from the Cosmic Microwave Background
radiation (CMB) \cite{WangW2013,HuangWW2015}. \\

Here we use the technique of minimising the likelihood, where we assume a Gaussian distribution for the free parameters:
\be
\mathcal{L}\propto \e^{-\chi^2/2}\ .
\label{likelihood}
\ee
The Pantheon SNe Ia  catalogue \cite{Pantheon17} includes $n_{\mbox{\scriptsize
SN}}=1048$ data points with redshifts $0< z_i\le2.26$ and their corresponding distance moduli
$\mu_i^{obs}$. Then, the theoretical models are compared with the data by calculating the theoretical value of the distance modulus  $\mu^{th}(z; \Omega_m^0, \lambda_i)$ for each set of the free parameters:
\be
 \mu^{th}(z; \Omega_m^0, \lambda_i) = 5 \log_{10} \frac{D_L(z; \Omega_m^0, \lambda_i)}{10\mbox{pc}} \ ,
 \label{mu} \end{equation}
 where $\lambda_i$ are the free parameters of the theoretical model and $D_L(z; \Omega_m^0, \lambda_i)$ is the free luminosity distance, which is given by:
\be
 D_L (z; \Omega_m^0, \lambda_i)= c (1+z) \int_0^z \frac{d\tilde z}{H (\tilde z)}\ .
 \label{lumdistance}
\ee
 Hence, $\chi^2$ function yields:
 \begin{equation}
\chi^2_{\mbox{\scriptsize SN}}(\Omega_m^0,A,\dots)=\min\limits_{H_0} \sum_{i,j=1}^{1048}
\Delta\mu_i\big(C_{\mbox{\scriptsize SN}}^{-1}\big)_{ij} \Delta\mu_j,\qquad
\Delta\mu_i=\mu^{th}(z_i,\Omega_m^0,\dots)-\mu^{obs}_i\ .
  \label{chiSN}\end{equation}
 Here  $C_{\mbox{\scriptsize SN}}$ is the $1048\times1048$ covariance matrix  \cite{Pantheon17}. For any set of the model parameters (\ref{6param}) we solve the system of equations provided in  (\ref{EqF}) and (\ref{Eqcont}), obtaining the Hubble parameter $H(z)$, and consequently the luminosity distances (\ref{lumdistance})
and the distance moduli (\ref{mu}). We also marginalise the  $\chi^2_{\mbox{\scriptsize SN}}$ function over the nuisance parameter $H_0$ \cite{OdintsovSGS:2017,Odintsov_Flog:2019,Sharov16,PanSh17}.\\



Baryonic Acoustic Oscillations (BAO) are provided by the analysis of galaxy clustering and the following two magnitudes can be compared with the observational data \cite{Eisen05}:
 \begin{equation}
 d_z(z)= \frac{r_s(z_d)}{D_V(z)},\qquad
  A(z) = \frac{H_0\sqrt{\Omega_m^0}}{cz}D_V(z)\ ,
  \label{dzAz} \end{equation}
where
$$ D_V(z)=\bigg[\frac{cz D_M^2(z)}{H(z)}\bigg]^{1/3},\qquad
D_M(z)=\frac{D_L(z)}{1+z}= c \int_0^z \frac{d\tilde z}{H (\tilde z)}\ ,$$
whereas  $r_s(z_d)$ is the comoving sound horizon at the end of the
baryon drag era $z_d$, which corresponds to a peak in the correlation function of the galaxy distribution.

As in previous works (see Refs.~\cite{OdintsovSGS:2017,Odintsov_Flog:2019}), here we use
 17 BAO data points for $d_z(z)$ and 7 data points for $A(z)$ from Refs.~\cite{BAOdata} estimated for galaxy clusters with mean redshifts $z=z_i$ and
represented in Table \ref{TBAO}.
\begin{table}[th]
\centering
 {\begin{tabular}{||l|l|l|l|l|l||}
\hline
 $z$  & $d_z(z)$ &$\sigma_d$    & $ A(z)$ & $\sigma_A$   & Survey\\ \hline
 0.106& 0.336  & 0.015 & 0.526& 0.028&  6dFGS \\ \hline
 0.15 & 0.2232 & 0.0084& -    & -    &  SDSS DR7  \\ \hline
 0.20 & 0.1905 & 0.0061& 0.488& 0.016&  SDSS DR7 \\ \hline
 0.275& 0.1390 & 0.0037& -    & -    &  SDSS DR7 \\ \hline
 0.278& 0.1394 & 0.0049& -    & -    & SDSS DR7 \\ \hline
 0.314& 0.1239 & 0.0033& -    & -    &  SDSS LRG \\ \hline
 0.32 & 0.1181 & 0.0026& -    & -    & BOSS DR11 \\ \hline
 0.35 & 0.1097 & 0.0036& 0.484& 0.016& SDSS DR7 \\ \hline
 0.35 & 0.1126 & 0.0022& -    & -    & SDSS DR7 \\ \hline
 0.35 & 0.1161 & 0.0146& -    & -    & SDSS DR7 \\ \hline
 0.44 & 0.0916 & 0.0071& 0.474& 0.034&  WiggleZ \\ \hline
 0.57 & 0.0739 & 0.0043& 0.436& 0.017&  SDSS DR9 \\ \hline
 0.57 & 0.0726 & 0.0014& -    & -    &  SDSS DR11 \\ \hline
 0.60 & 0.0726 & 0.0034& 0.442& 0.020&  WiggleZ \\ \hline
 0.73 & 0.0592 & 0.0032& 0.424& 0.021& WiggleZ \\ \hline
 2.34 & 0.0320 & 0.0021& -& - &  BOSS DR11 \\ \hline
 2.36 & 0.0329 & 0.0017& -& - &  BOSS DR11 \\  \hline
 \end{tabular}
 \caption{BAO data $d_z(z)=r_s(z_d)/D_V(z)$ and $A(z)$ (\ref{dzAz}).}
 \label{TBAO}} \end{table}

For  the sound horizon $r_s(z_d)$ we use the fitting formula \cite{OdintsovSGS:2017,Odintsov_Flog:2019,Sharov16}
\be
 r_s(z_d)=\frac{104.57\mbox{ Mpc}}h,\qquad
 h=\frac{H_0}{100\mbox{ km}/(\mbox{s}\cdot\mbox{Mpc})}\, ,
 \label{soundhorizon}
 \ee
 which shows a dependence on the Hubble parameter $r_s(z_d)\sim H_0^{-1}$ and leaves  $d_z(z)$ Hubble free.
Then, the $\chi^2$ function for the BAO fits (\ref{dzAz}) is
 \begin{equation}
 \chi^2_{\mbox{\scriptsize BAO}}(\Omega_m^0,A,\dots)=\Delta d\cdot C_d^{-1}(\Delta d)^T+
\Delta { A}\cdot C_A^{-1}(\Delta { A})^T,
  \label{chiB} \end{equation}
 where $\Delta d$, $\Delta A$ are vector columns with elements
 $\Delta d_i=d_z^{obs}(z_i)-d_z^{th}(z_i,\dots)$; $\Delta A_i=A^{obs}(z_i)-A^{th}(z_i,\dots)$, $C_{d}$ and $C_{A}$ are the covariance matrices for correlated BAO data \cite{BAOdata} described in Ref.~\cite{Sharov16}.\\

In addition, the Hubble parameter $H(z)$ data are given by $N_H=31$ data points
$H^{obs}(z_i)$ from Refs.~\cite{Hdata} for redshifts $0<z<2 $, whose $\chi^2$ function
yields: $\chi^2$ function
 \begin{equation}
\chi^2_{H}=\min\limits_{H_0} \sum_{i=1}^{N_H} \left[\frac{H^{obs}(z_i)-H^{th}(z_i,
\alpha,\dots)}{\sigma_{H,i}}\right]^2 \ .
 \label{chiH} \end{equation}
Here we use only data \cite{Hdata} estimated by the method of differential ages (cosmic chronometers),  where the values for the Hubble parameter at different redshifts
 $$ 
 H (z)= \frac{\dot{a}}{a}= -\frac{1}{1+z}\frac{dz}{dt}  \simeq -\frac{1}{1+z}
\frac{\Delta z}{\Delta t}\ ,
 $$
are estimated at differential ages $\Delta t$ of galaxy clusters with certain differences $\Delta z$
of redshifts. These estimations are not correlated with the BAO data points \cite{BAOdata} at the level (\ref{chiB}).\\

Finally, we will also use the CMB parameters for testing our models. Unlike the datasets coming from SNe Ia, BAO and $H(z)$ observations, measured for
$0<z\le2.36$, the CMB observational parameters are related with the photon-decoupling epoch $z_*=1089.90 \pm0.25$ \cite{Planck13,Planck18}. Hence, as we are dealing with high redshifts here, the radiation density is not negligible and enters in the equations through the radiation-matter ratio $X_r=\rho_r^0/\rho_m^0$  in the form (\ref{Xrm}). We use the CMB parameters released by Planck \cite{Planck13,Planck15} in the following form \cite{WangW2013,HuangWW2015}:
 \begin{equation}
  \mathbf{x}=\big(R,\ell_A,\omega_b\big);\qquad R=\sqrt{\Omega_m^0}\frac{H_0D_M(z_*)}c,\quad
 \ell_A=\frac{\pi D_M(z_*)}{r_s(z_*)},\quad\omega_b=\Omega_b^0h^2,
 \label{CMB} \end{equation}
 where  the comoving sound horizon $r_s$ at $z_*$ is calculated as
  $$
  r_s(z)=\frac1{\sqrt{3}}\int_0^{1/(1+z)}\frac{da}
 {a^2H(a)\sqrt{1+\big[3\Omega_b^0/(4\Omega_r^0)\big]a}}\ .
   $$
The current baryon fraction $\Omega_b^0$ is considered as the nuisance parameter and it is
marginalized over $\omega_b=\Omega_b^0h^2$ and $H_0$ in the $\chi^2_{CMB}$ function
 \begin{equation}
\chi^2_{\mbox{\scriptsize CMB}}=\min_{H_0,\omega_b}\Delta\mathbf{x}\cdot
C_{\mbox{\scriptsize CMB}}^{-1}\big(\Delta\mathbf{x}\big)^{T},\qquad \Delta
\mathbf{x}=\mathbf{x}-\mathbf{x}^{Pl}\ .
 \label{chiCMB} \end{equation}
  The following data are provided in \cite{HuangWW2015} from Planck collaboration \cite{Planck15}:
  \begin{equation}
  \mathbf{x}^{Pl}=\big(R^{Pl},\ell_A^{Pl},\omega_b^{Pl}\big)=\big(1.7448\pm0.0054,\;301.46\pm0.094,\;0.0224\pm0.00017\big)
   \label{CMBpriors} \end{equation}
which are given with free amplitude for the lensing power spectrum. The covariance matrix $C_{\mbox{\scriptsize
CMB}}=\|\tilde C_{ij}\sigma_i\sigma_j\|$,  $\tilde C_{12}=0.53$,  $\tilde C_{13}=-0.73$,
$\tilde C_{23}=-0.42$ and other details are described in \cite{HuangWW2015} and also in Refs.~\cite{OdintsovSGS:2017,Odintsov_Flog:2019}.

\section{Results and discussion}
\label{Results}

Here the above SNe Ia, BAO,  $H(z)$ and CMB datasets are used to constrain the models (\ref{EoS1}) and (\ref{EoSln1}) described in section \ref{Dynamics}, through the analysis of the parameter space (\ref{6param}) in order to obtain the best fit and the confidence regions for each of the free parameters. The CMB
observations (\ref{chiCMB}) with small errors $\sigma_i$ (\ref{CMBpriors})  produce the
most strict limitations in the  parameter space in comparison with other data. For that reason,
we analyse separately the $\chi^2$ function obtained after fitting the free parameters with SNe Ia, BAO and  $H(z)$ data at redshifts $0<z\le2.36$ \cite{OdintsovSGS:2017,Odintsov_Flog:2019}:
 \begin{equation}
  \chi^2_{\Sigma3}=\chi^2_{\mbox{\scriptsize SN}}+\chi^2_H+\chi^2_{\mbox{\scriptsize BAO}}
 \label{chi3} \end{equation}
Whereas we estimate the total $\chi^2_{tot}$ separately:
 \begin{equation}
  \chi^2_{\mbox{\scriptsize tot}}=\chi^2_{\mbox{\scriptsize SN}}+\chi^2_H+\chi^2_{\mbox{\scriptsize BAO}}+\chi^2_{\mbox{\scriptsize CMB}},
 \label{chitot} \end{equation}
 where $\chi^2_{\mbox{\scriptsize CMB}}$ corresponds to redshifts near $ z_*\simeq1100$.\\

\begin{figure}[th]
   \centerline{ \includegraphics[scale=0.66,trim=5mm 0mm 2mm -1mm]{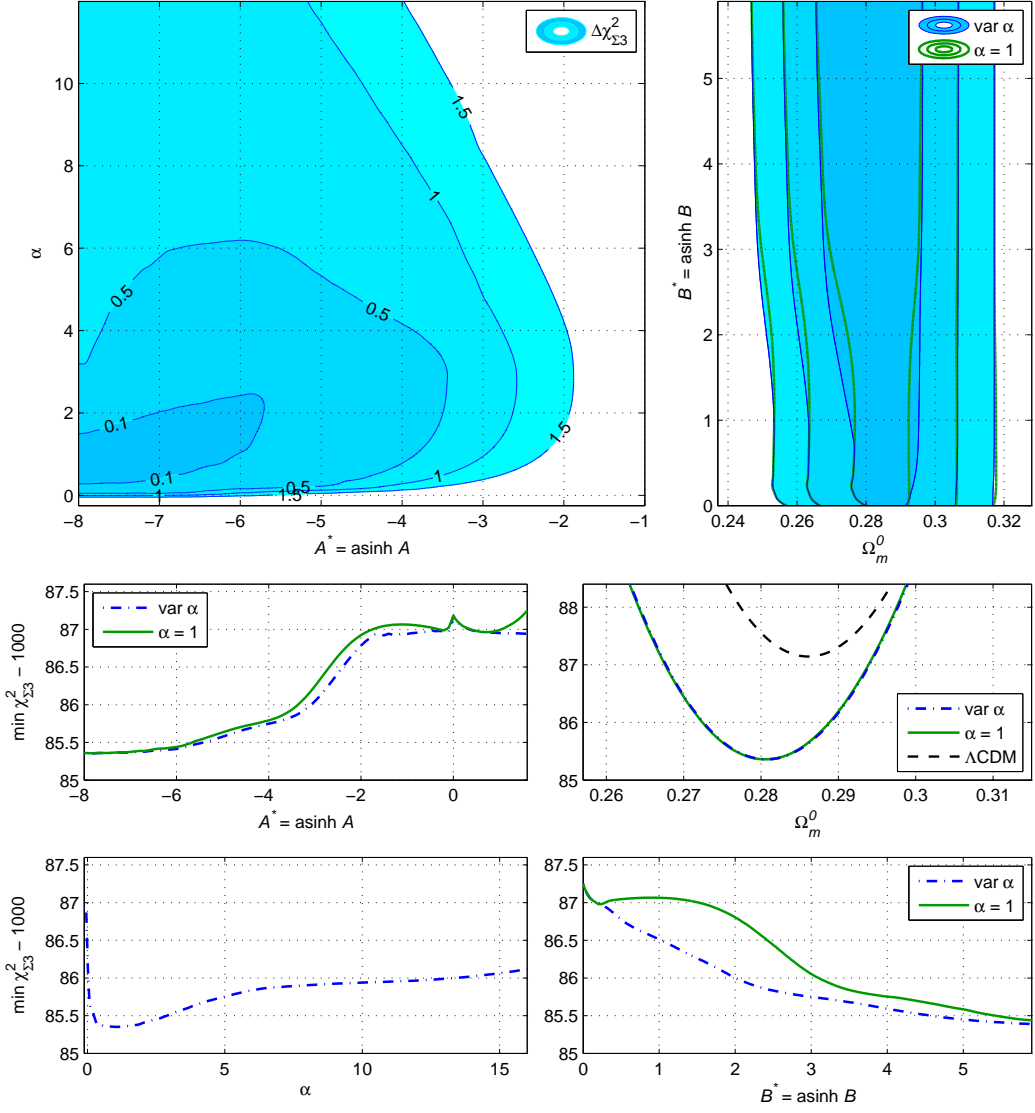}}
\caption{Contours plots and statistical distributions for the power-law model (\ref{EoS1}). Top panels show the contour plots filled in blue for the corresponding confidence regions: the top-left panel depicts the $A-\alpha$ plane, where the blue regions indicates the values of $\Delta\chi^2_{\Sigma3}=\chi^2_{\Sigma3}(A,\alpha)-m^{\mbox{\scriptsize abs}}_{\Sigma3}$, which are properly indicated. In the top-right panel, the $1\sigma$, $2\sigma$ and $3\sigma$ contours for $\chi^2_{\Sigma3}$
are shown in the $\Omega_m^0-B$ plane. Green lines refer to the contour plots when setting $\alpha=1$. In the bottom panels we present the corresponding one-parameter distributions
 $\min\chi^2_{\Sigma3}$ for the parameters $\alpha$, $A$, $\Omega_m^0$ and $B$, including the case of varying $\alpha$
(blue dash-dotted lines) and the fixed one case $\alpha=1$ (\ref{alp1}) (green lines). $\Lambda$CDM model is also depicted, represented by the black dashed line at the
 $\Omega_m^0$ plot).
 }
  \label{F1}
\end{figure}

Let us start by calculating the $\chi^2_{\Sigma3}$ function (\ref{chi3}) for
the power-law model (\ref{EoS1}), with the free parameters as given in (\ref{6param}). The results are shown in Fig.~\ref{F1}, particularly the
$A-\alpha$ contour plot is depicted in the top-left panel, where we have minimised the $\chi^2$ function over the other parameters and have calculated the difference among the absolute minimum and its variation as a function of $A-\alpha$:
 $$ 
\Delta\chi^2_{\Sigma3}(A,\alpha)=\min\limits_{\Omega_m^0,B,\beta}\chi^2_{\Sigma3}-m^{\mbox{\scriptsize
abs}}_{\Sigma3}.
 $$ 
 Here $m^{\mbox{\scriptsize
abs}}_{\Sigma3}=\min\limits_{\mbox{\scriptsize all}}\chi^2_{\Sigma3}$ is the absolute
minimum of
 $\chi^2_{\Sigma3}$ over all its parameters  $A$, $\alpha$, $\Omega_m^0$, $B$, $\beta$
which in this case is  $m^{\mbox{\scriptsize abs}}_{\Sigma3}\simeq1085.35$.
In the top-left panel of Fig.~\ref{F1} we depict the two-parameter distribution
$\Delta\chi^2_{\Sigma3}(A,\alpha)$, where blue lines represents the values of $\Delta\chi^2_{\Sigma3}$, as indicated in the figure. The dependence of $\min\chi^2_{\Sigma3}$ on $A$ and $\alpha$ is very weak and the whole depicted area in the
$A-\alpha$ plane lies in the $1\sigma$ confidence region. At the bottom panels, this weak
dependence is also shown for the corresponding one-parameter distributions, where
 \begin{equation}
\min\chi^2_{\Sigma3}(\alpha)\equiv
\chi^2_{\Sigma3}(\alpha)=\min\limits_{A,\Omega_m^0,B,\beta}\chi^2_{\Sigma3}\,,
 \label{minalp} \end{equation}
is minimised over all the other parameters.\\

As shown in Fig.~\ref{F1}, the minimum value for $\Delta\chi^2_{\Sigma3}$ in the $A-\alpha$ plane lies within the area with large
negative values of A and $ \alpha\sim1$ (see the top-left panel), so for convenience, we can fix the value for $\alpha$ without loss of information and effectiveness when minimising $\chi^2$ for this model,
 \begin{equation}
 \alpha=1\ .
 \label{alp1} \end{equation}
Indeed, the minimum for $\chi^2_{\Sigma3}$ under the assumption (\ref{alp1}) is only a bit larger than the
absolute minimum:
 $$
 \min\limits_{\Omega_m^0,A,B,\beta}\chi^2_{\Sigma3}\Big|_{\alpha=1}\simeq 1085.36\;,\qquad
 m^{\mbox{\scriptsize
abs}}_{\Sigma3}\equiv
\min\limits_{\Omega_m^0,A,\alpha,B,\beta}\chi^2_{\Sigma3}\simeq1085.35\;.
 $$
By fixing the value for  $\alpha$ as given in (\ref{alp1}), the remaining free parameters are (recall we have marginalise over $H_0$):
 \begin{equation}
\Omega_m^0, \; A,\; B,\;\beta\,,
 \label{4param}
\end{equation}
 and its EoS (\ref{EoS1}) is reduced to
 \begin{equation}
\frac{p_x}{\rho_{\mbox{\scriptsize cr}}} =(A-1)\,\Omega_x +
B\bigg(\frac{H}{H_0}\bigg)^{2\beta}.
 \label{EoSa1}\end{equation}
The top-right panel of Fig.~\ref{F1} shows the $\Omega_m^0-B$ plane of the
two-parameter distribution $\min\chi^2_{\Sigma3}(\Omega_m^0,B)$ (minimised over the other
parameters) for the model (\ref{EoS1}) for a varying $\alpha$  (blue filled contours) and
for the restricted case $\alpha=1$ (green contours). One
can see that the contours of $1\sigma$, $2\sigma$ and $3\sigma$ confidence regions for
both cases do not differ much. This similarity is the most striking, when we compare the one-parameter distributions
$\min\limits_{other}\chi^2_{\Sigma3}(\Omega_m^0)$ in the panel below: the corresponding
green and blue dash-dotted lines practically coincide. The $\Lambda$CDM model is also depicted for comparison with our model (the black dashed line). Note that the power-law model (\ref{EoS1}) transforms into the $\Lambda$CDM model for $\Omega_x=\mbox{const}=\Omega_\Lambda$, which corresponds to the particular case $A=B=0$. In addition, the difference among the cases $\alpha=1$ and $\alpha\in R$ becomes more remarkable in the
bottom-right panel, where $\min\chi^2_{\Sigma3}$ depends on $B$. As shown in Fig.~\ref{F1} and Table \ref{Estim}, both parameters $\{A, B\}$ are unbounded, since the function $\chi^2_{\Sigma3}$ extends its 1$\sigma$ region up to $A\to-\infty$ and
$B\to+\infty$, respectively. \\

\begin{figure}[th]
   \centerline{ \includegraphics[scale=0.66,trim=5mm 0mm 2mm -1mm]{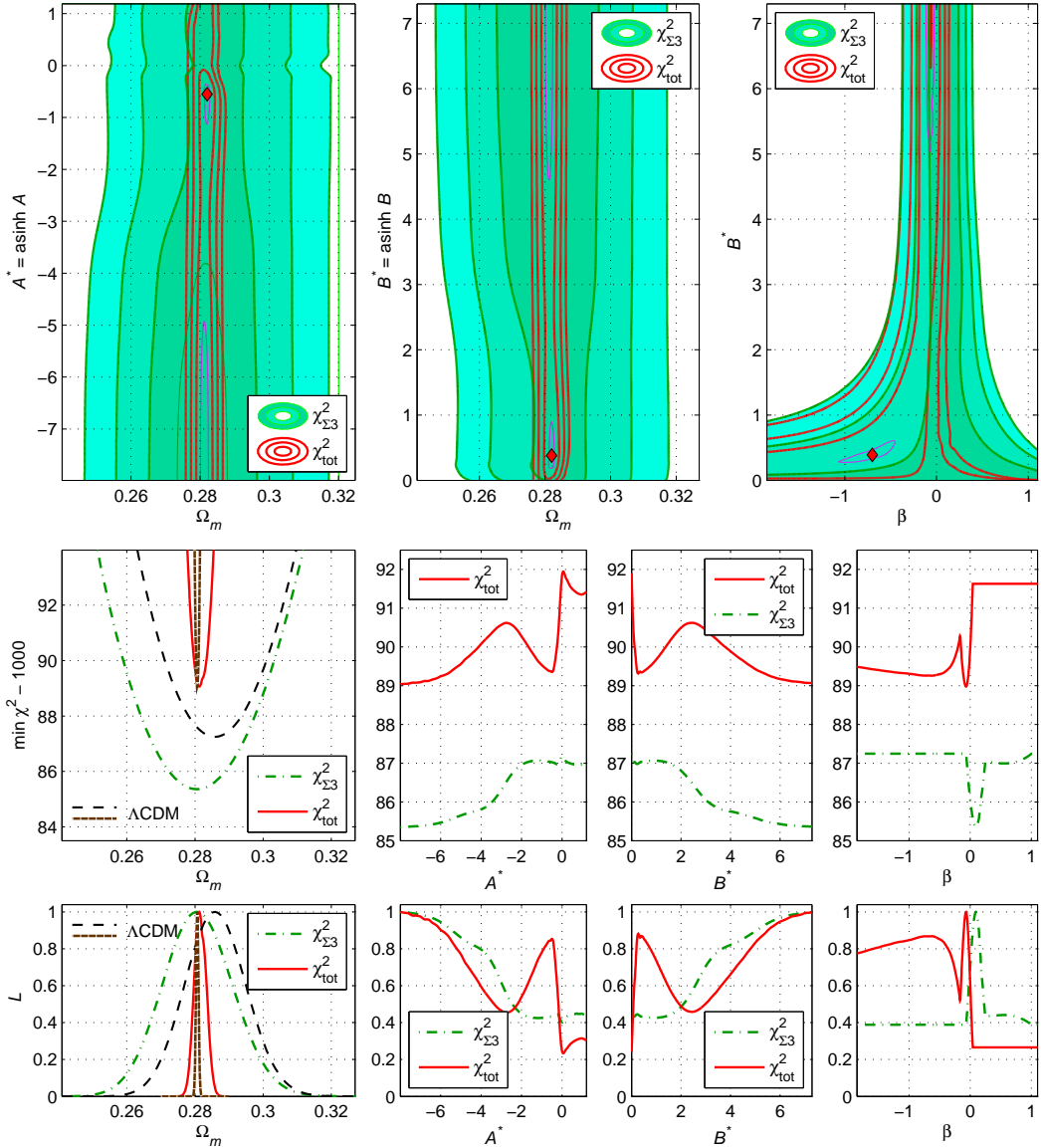}}
\caption{Counter plots and statistical distributions for the power-law model (\ref{EoSa1}) when assuming $\alpha=1$. The top panels depict the confidence regions for $\Omega_m^0-A^*$,
$\Omega_m^0-B^*$ and $\beta-B^*$, with each region corresponding to the $1\sigma$, $2\sigma$ and
$3\sigma$.  The green filled contours areas refer to $\chi^2_{\Sigma3}$ and red ones to the total
$\chi^2_{\mbox{\scriptsize tot}}$. Middle panels 
provide the minimum $\chi^2$, also for both sets of fits, $\chi^2_{\Sigma3}(p_j)$,
$\chi^2_{\mbox{\scriptsize tot}}(p_j)$, while the bottom panels are the likelihood functions
 ${\cal L}_{\Sigma3}(p_j)$, ${\cal L}_{\mbox{\scriptsize tot}}(p_j)$, where
 $p_j=\Omega_m^0$, $A$, $B$ and $\beta$. The one for  $\Omega_m^0$ includes the statistical distribution for the $\Lambda$CDM model.
 }
  \label{F2}
\end{figure}

Hence, the model (\ref{EoSa1}) with $\alpha=1$ provides effectively
the same results as the general model (\ref{EoS1}) with $\alpha\in R$, since there is no correlation among $\alpha$ and the other parameters, as shown in Fig.~\ref{F1}. From here on, we assume $\alpha=1$. In addition, motivated by the behaviour of $\chi^2_{\Sigma3}$ in the parameter space, we redefine the parameters as $A^*$ and $B^*$, which are related to $A$ and $B$ by:
 \begin{equation}
A=\sinh A^*,\qquad B=\sinh B^*.
 \label{ABsh}\end{equation}
In Fig.~\ref{F2} we investigate in detail the EoS (\ref{EoSa1}) with $\alpha=1$ in
the $\Omega_m^0-A^*$, $\Omega_m^0-B^*$ and $\beta-B^*$ planes including the CMB data
(\ref{chiCMB}), (\ref{CMBpriors}): the corresponding $1\sigma$, $2\sigma$, $3\sigma$
contour plots (top panels) are depicted for the function (\ref{chitot})
$\chi^2_{\mbox{\scriptsize tot}}=\chi^2_{\Sigma3}+\chi^2_{\mbox{\scriptsize CMB}}$  by red lines, the red diamonds show the local minimum points of
$\chi^2_{\mbox{\scriptsize tot}}$. Green filled contours and green dash-dotted lines in
all panels correspond to the function $\chi^2_{\Sigma3}$ (for the $\Omega_m^0-B^*$ plane these contours were
shown in Fig.~\ref{F1}).

As mentioned above, the CMB observational data, given in (\ref{chiCMB}) and (\ref{CMBpriors}), impose the most severe constraints, specially for the parameter $\Omega_m^0$, as obtained after analysing the corresponding $\chi^2_{\mbox{\scriptsize
tot}}=\chi^2_{\Sigma3}+\chi^2_{\mbox{\scriptsize CMB}}$ (see Fig.~\ref{F2}). This is
connected to the observational data $R=1.7448\pm0.0054$, which is proportional to $\sqrt{\Omega_m^0}$. One can see that the
function (\ref{chitot})  has the additional local minimum at $\Omega_m^0\simeq0.282$, $A^*\simeq-0.55$, $B^*\simeq0.38$, but is not the
global minimum as shown in Fig.~\ref{F2}).

In the middle row of Fig.~\ref{F2}, the one-parameter distributions of the
type $\chi^2_{\Sigma3}(p_j)$ and  $\chi^2_{\mbox{\scriptsize tot}}(p_j)$ are depicted 
for the 4 parameters $p_j=\Omega_m^0$, $A^*$, $B^*$ and $\beta$. The corresponding
likelihoods ${\cal L}_{\Sigma3}(p_j)$ and ${\cal L}_{\mbox{\scriptsize tot}}(p_j)$ are
shown in the bottom panels. These functions are obtained for  ${\cal
L}_{\mbox{\scriptsize tot}}$ as follows:
  $$ 
\chi^2_{\mbox{\scriptsize tot}}(p_j)=\min\limits_{\mbox{\scriptsize other
}p_k}\chi^2_{\mbox{\scriptsize tot}}(p_1,\dots),\qquad {\cal L}_{\mbox{\scriptsize
tot}}(p_j)= \exp\bigg[- \frac{\chi^2_{\mbox{\scriptsize tot}}(p_j)-m_{\mbox{\scriptsize
tot}}^{\mbox{\scriptsize abs}}}2\bigg]\ ,
 $$ 
 where the function is marginalised over all the other free parameters, being $m_{\mbox{\scriptsize tot}}^{\mbox{\scriptsize abs}}$
the absolute minimum for $\chi^2_{\mbox{\scriptsize tot}}$.\\

In the middle-left and bottom-left panels we compare these results with the
corresponding  distributions for the $\Lambda$CDM model for  $\chi^2_{\Sigma3}(\Omega_m^0)$, ${\cal L}_{\Sigma3}(\Omega_m^0)$ (black
dashed lines) and for $\chi^2_{\mbox{\scriptsize tot}}(\Omega_m^0)$,  ${\cal
L}_{\mbox{\scriptsize tot}}(\Omega_m^0)$ (brown lines). One can see that for the model
(\ref{EoSa1}) with $\alpha=1$, the absolute minimum for $\chi^2_{\Sigma3}$ is essentially
lower than the corresponding value for the $\Lambda$CDM model, but it is not true for
$\chi^2_{\mbox{\scriptsize tot}}$, when including the CMB data. In addition, we should also mention the the optimal values for $A$ and
$B$ go to $-\infty$ and  $+\infty$ respectively. Note that the domain $B>0$ corresponds to negative viscosity $\zeta$ in Eq.~(\ref{EoSvisc}).

The values for the absolute minimum and the best fits (with $1\sigma$ errors) of the free model parameters for the model (\ref{EoSa1}) are
given in Table~\ref{Estim}. The results of the logarithmic model  (\ref{EoSln1}) are also included. The best fits and
$1\sigma$ errors are calculated via the distributions $\chi^2(p_j)$ or ${\cal L}(p_j)$.

 \begin{figure}[th]
   \centerline{ \includegraphics[scale=0.66,trim=5mm 0mm 2mm -1mm]{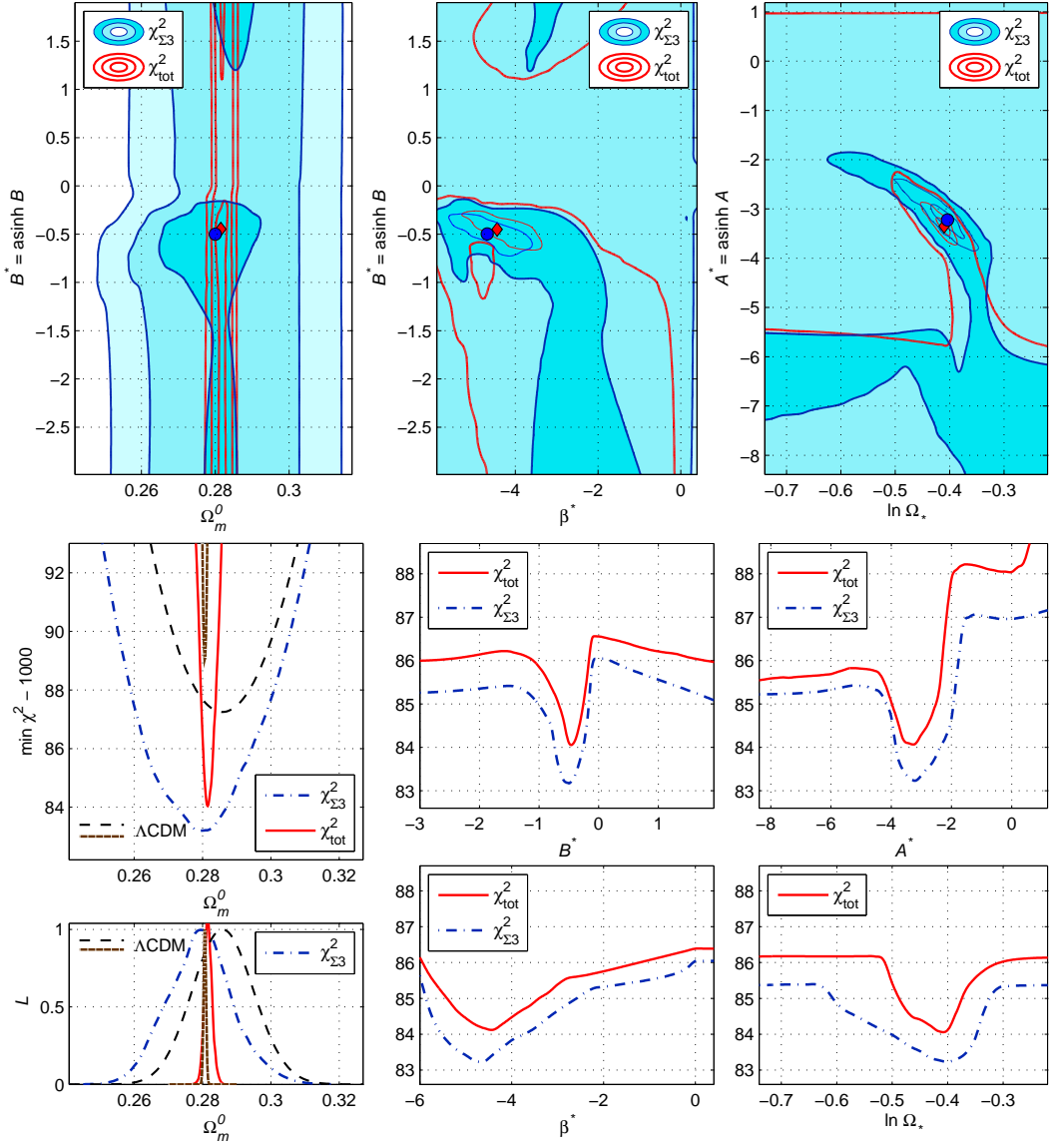}}
\caption{Contour plots for the logarithmic model  (\ref{EoSln1}). As above, we depict the confidence regions when considering $\chi^2_{\Sigma3}$
(filled contours) and for $\chi^2_{\mbox{\scriptsize tot}}$ (red lines). In the bottom panels, the distributions for
 $\Omega_m-B^*$, $\beta^*-B^*$ and $\Omega_*-A^*$  are also depicted.
 }
  \label{F3}
\end{figure}

\medskip

On the other hand, the logarithmic model  (\ref{EoSln1}) behaves in another way and shows better
results (see Table~\ref{Estim} and in Fig.~\ref{F3}). The minimums for $\chi^2_{\Sigma3}$ and $\chi^2_{\mbox{\scriptsize tot}}$ are essentially
smaller than in the $\Lambda$CDM model and the power-law case. Indeed, in Table~\ref{Estim} we
can compare, for example, $\min\chi^2_{\mbox{\scriptsize tot}}\simeq1084.05$ for the
logarithmic model with the corresponding  $\Lambda$CDM minimum $1089.03$. Moreover, unlike the power-law scenario, these minimums in the logarithmic model (\ref{EoSln1}) are
achieved at finite values of $A$ and $B$ with $B<0$, corresponding to positive viscosity $\zeta$.\\

Fig.~\ref{F3} shows in detail the contour plots for the $\Omega_m-B^*$,
$\beta^*-B^*$, $\Omega_*-A^*$  planes  and the one-parameter distributions for the
logarithmic model (\ref{EoSln1}). The points of minimum are labeled as blue circles for
$\chi^2_{\Sigma3}$ and as red diamonds for $\chi^2_{\mbox{\scriptsize tot}}$. These
functions behave non-trivially in some domains of the parameter space as can be seen
in Fig.~\ref{F3}. In particular, in the contour plots for $\beta^*-B^*$ and for
$\Omega_*-A^*$ (the top-center and top-right panels) the borders of
the $1\sigma$ and $2\sigma$ confidence region are not regular, whereas the $3\sigma$ domain lies beyond.
These unusual behaviour can be also seen in the corresponding one-parameter distributions
$\min\limits_{other}\chi^2(p_i)$ of Fig.~\ref{F3}, particularly these functions decrease at large
negative values of $B^*$ and $A^*$ and positive values of $B^*$. However, here the local
minimum coincide with the global minimum for the $\chi^2$ function, unlike the power-law model.

 \begin{table}[ht]
 \centering
 {\begin{tabular}{||l|c||c|c|c|c|c|l||}  \hline
 Model  & data & $\Omega_m^0$& $A^*$ & $B^*$ & $\beta^*$ &$\Omega_*$ &$ \min\chi^2/d.o.f$  \\ \hline\hline
power-law & $\chi^2_{\Sigma3}$& $0.2808_{-0.0102}^{+0.0104}$ & $-8.8_{-\infty}^{+5.85}$  & 
$7.94_{-5.34}^{+\infty}$  & $0.092_{-0.112}^{+0.108}$ & - & 1085.36 / 1099\rule{0pt}{1.2em}  \\
 \hline
logarithmic& $\chi^2_{\Sigma3}$& $0.280_{-0.009}^{+0.008}$ & $-3.22_{-0.66}^{+0.99}$  & 
$-0.50_{-0.26}^{+0.265}$  & $-4.68_{-0.95}^{+1.24}$ &  $-0.405_{-0.125}^{+0.062}$ & 1083.20 / 1098\rule{0pt}{1.2em}  \\
 \hline
$\Lambda$CDM & $\chi^2_{\Sigma3}$& $0.286^{+0.0089}_{-0.009}$ & - & - & - & - & 1087.25 / 1102 \rule{0pt}{1.2em} \\
 \hline \hline
 power-law& $\chi^2_{\mbox{\scriptsize tot}}$& $0.2815_{-0.0018}^{+0.0019}$ & $-9.2_{-\infty}^{+5.04}$ &
$8.15_{-4.75}^{+\infty}$  & $-0.068_{-0.082}^{+0.068}$ & - & 1088.98 / 1102 \rule{0pt}{1.2em}  \\
 \hline
 logarithmic & $\chi^2_{\mbox{\scriptsize tot}}$& $0.2815_{-0.0009}^{+0.0012}$ & $-3.35_{-0.65}^{+0.84}$ &
$-0.45_{-0.235}^{+0.198}$  & $-4.44_{-0.89}^{+1.16}$  & $-0.41_{-0.08}^{+0.037}$ & 1084.05 / 1101 \rule{0pt}{1.2em}  \\
 \hline
$\Lambda$CDM & $\chi^2_{\mbox{\scriptsize tot}}$& $\!0.2807_{-0.0004}^{+0.0003}\!$ & - & - & - & - & 1089.03 / 1105 \rule{0pt}{1.2em} \\
 \hline \end{tabular}
\caption{Best fits for the power-law model (\ref{EoSa1}) with $\alpha=1$ and the
logarithmic model  (\ref{EoSln1}), when considering $\chi^2_{\Sigma3}=\chi^2_{\mbox{\scriptsize SN}}+\chi^2_{\mbox{\scriptsize
BAO}}+\chi^2_H$ (the Pantheon SNe Ia, BAO and $H(z)$ data) and for $\chi^2_{\mbox{\scriptsize tot}}=\chi^2_{\Sigma3}+\chi^2_{\mbox{\scriptsize CMB}}$
 (including the CMB data) in comparison with the flat $\Lambda$CDM model. The table also shows the $\min\chi^2$ and the $1\sigma$ errors for the model parameters. Here $\beta=\sinh(\beta^*)$, similar to the relations (\ref{ABsh}).}
 \label{Estim}}
 \end{table}
Note that the narrow peak of $\chi^2_{\mbox{\scriptsize tot}}$ for the $\Lambda$CDM
model in the bottom-left panels of Figs.~\ref{F2} and \ref{F3} is connected with the
 CMB parameter $R\sim\sqrt{\Omega_m^0}$ in Eqs.~(\ref{chiCMB}) and
(\ref{CMBpriors}), and with the number of free parameters for the flat $\Lambda$CDM
model, that is $\Omega_m^0$ and the nuisance parameter $H_0$, as shown in the FLRW equation:
 $$
 H^2\big/H_0^2=\Omega_m^0 a^{-3}+1-\Omega_m^0.
 $$
Nevertheless, the logarithmic model shows a good behaviour in comparison to $\Lambda$CDM, as shown in Fig.~\ref{F3} and Table \ref{Estim}.
\section{Conclusions}
\label{Conclusion}

Along the paper we have considered two cosmological scenarios where dark energy is assumed to be described by a viscous fluid, through bulk viscosity, what leads to an effective pressure that can explain the late-time accelerating expansion. For that, and inspired on some hydrodynamics considerations, we have explored two different EoS for 
viscous dark energy: the power-law model (\ref{EoS1}), precisely, its variant (\ref{EoSa1})
with $\alpha=1$ and the logarithmic model (\ref{EoSln1}). By using data from Supernovae Ia, BAO, $H(z)$ measurements and CMB, we have analysed the viability of these scenarios and compared to $\Lambda$CDM model.\\

Our analysis shows that the power-law model (\ref{EoSa1}) behaves well, also in comparison with $\Lambda$CDM model, when considering the restricted set of observational data that excludes CMB data, as depicted in Figs.~\ref{F1} and \ref{F2} and summarised in Table~\ref{Estim}). Actually, the model (\ref{EoSa1}) provides a slightly lower minimum when considering $\chi^2_{\Sigma3}$ than $\Lambda$CDM model, but higher errors for $\Omega_m$, and weak constraints on the free parameters $\{A, B\}$, since they show no bound from above/below, which may lead to a negative viscosity, as given in (\ref{EoSvisc}), particularly the model (\ref{EoSa1}) achieves the best $\chi^2$ values at the non-physical domain
$B\to+\infty$ that corresponds to large negative viscosity $\zeta$ in Eq.~(\ref{EoSvisc}). The other free parameter of the model $\alpha$ is very well constrained at $\alpha\sim 1$ and shows no correlations with the other free parameters, such that we have assumed $\alpha=1$ for a great part of our calculations, as depicted in Figs.~\ref{F1} and \ref{F2}. However, when assuming CMB data, the model (\ref{EoSa1}) provides larger values for the minimum of $\chi^2_{tot}$, but similar to the one given by $\Lambda$CDM model. This means that the viscous term as a power law behaves well at late-times, but shows some issues when increasing the covered region of the cosmological evolution. In addition, the same problems with the parameters $\{A, B\}$ remain in this case (see Table \ref{Estim}), and one can not obtain better constraints for both. \\

Unlike the power-law model  (\ref{EoSa1}), the logarithmic model (\ref{EoSln1}) has no these drawbacks: it provides essentially lower values for $\min\chi^2_{\Sigma3}$ and
$\min\chi^2_{\mbox{\scriptsize tot}}$, which are achieved at reasonable values of the free parameters, and the constraints on each parameter are well defined and limited (see Table~\ref{Estim}). The values of the minimums for $\chi^2$ show that the model (\ref{EoSln1}) fits better every set of observational data, in comparison to the power-law model and the $\Lambda$CDM model. However, despite the contour plots and statistical distributions in Fig.~\ref{F3} show a well defined 1$\sigma$ region, the errors increase much when one goes to confidence regions of upper $\sigma$, which may be interpreted as some lack of information on the free parameters. For instance, the analysis of $\chi^2_{\mbox{\scriptsize tot}}$ provides the best value for $B$ as $B=\sinh B^*=-0.465_{-0.275}^{+0.21}$, which corresponds to positive optimal values of viscosity $\zeta$,  strongly depending on $H$ because for negative $\beta$, but if one increases the confidence region, $B$ may take values that lead to a negative viscosity, and unconstrained model. In any case, the logarithmic model (\ref{EoSln1}) seems to provide a very well description of the cosmological evolution at any redshifts, that is also when including CMB data.\\

Hence, we have explored the existence of a viscous dark energy fluid by using the last observational data coming from different sources and by considering some theoretical models for the viscosity terms that play a role in other areas of hydrodynamics. Our results show that the right viscosity term can provide better fits in comparison to other models, such that one should keep analysing this possibility by going beyond the analysis of the cosmological background evolution.

\section*{Acknowledgments}
SDO acknowledges the support of MINECO (Spain), project FIS2016-76363-P, and AGAUR (Catalonia, Spain) project 2017 SGR 247. DS-CG is funded by the University of Valladolid (Spain). This article is based upon work from CANTATA COST (European Cooperation in Science and Technology) action CA15117,  EU Framework Programme Horizon 2020.

\end{document}